\documentclass[10pt,conference]{ieeeconf}

\usepackage{mathrsfs}
\usepackage{graphics} 
\usepackage{amsmath} 
\usepackage{amssymb}  

\usepackage{amsthm}

\usepackage{cite}
\usepackage{bm}
\usepackage{acronym}
\usepackage{paralist}
\usepackage{float}
\usepackage{color}
\usepackage{epstopdf}
\usepackage{multicol}
\usepackage{tikz}
\usepackage{graphicx}
\usepackage{mathtools}
\usepackage{soul}
\usepackage[hang, flushmargin]{footmisc}
\usepackage[colorlinks=true]{hyperref}

\usepackage{amsmath} 
\usepackage{amssymb}  
\usepackage{graphicx}
\usepackage{epstopdf}
\usepackage{amsmath}
\usepackage{mathtools}
\usepackage{dsfont}
\usepackage{tikz}
\usepackage{siunitx}
\usepackage{xcolor}

\newcommand{\pn}[1]{{\color{black} #1}}
\newcommand{\pno}[1]{{\color{black} #1}}

\usepackage{ifthen}
\newboolean{Disturbances}
\setboolean{Disturbances}{false}
\newcommand{\NotDist}[1]{\ifthenelse{\boolean{Disturbances}}{}{{\color{black}#1}}}
\newcommand{\IfDist}[1]{\ifthenelse{\boolean{Disturbances}}{\color{red}#1}{}}  
\newboolean{ACCFinal}
\setboolean{ACCFinal}{True}
\newcommand{\NotACC}[1]{\ifthenelse{\boolean{ACCFinal}}{}{{\color{magenta}#1}}}
\newcommand{\IfACC}[2]{\ifthenelse{\boolean{ACCFinal}}{{\color{black}#1}}{{\color{magenta}#2}}}  

\newboolean{Integrity}
\setboolean{Integrity}{false}
\newcommand{\NotInt}[1]{\ifthenelse{\boolean{Integrity}}{}{{\color{black}#1}}}
\newcommand{\IfInt}[1]{\ifthenelse{\boolean{Integrity}}{{\color{red}#1}}}

\makeatletter
\hypersetup{colorlinks=true}
\AtBeginDocument{\@ifpackageloaded{hyperref}
  {\def\@linkcolor{blue}
  \def\@anchorcolor{red}
  \def\@citecolor{red}
  \def\@filecolor{red}
  \def\@urlcolor{black}
  \def\@menucolor{red}
  \def\@pagecolor{red}
\begingroup
  \@makeother\`%
  \@makeother\=%
  \edef\x{%
    \edef\noexpand\x{%
      \endgroup
      \noexpand\toks@{%
        \catcode 96=\noexpand\the\catcode`\noexpand\`\relax
        \catcode 61=\noexpand\the\catcode`\noexpand\=\relax
      }%
    }%
    \noexpand\x
  }%
\x
\@makeother\`
\@makeother\=
}{}}
\makeatother

\newtheorem{Theorem}{Theorem}

\newtheorem*{Problem}{Problem $(\star)$}

\newtheorem{Remark}{Remark}

\newtheorem{Lemma}{Lemma}

\newtheorem{Assumption}{Assumption}

\newtheorem{Definition}{Definition}

\IEEEoverridecommandlockouts

\def\BibTeX{{\rm B\kern-.05em{\sc i\kern-.025em b}\kern-.08em
    T\kern-.1667em\lower.7ex\hbox{E}\kern-.125emX}}

\makeatletter
\newcommand{\fixed@sra}{$\vrule height 2\fontdimen22\textfont2 width 0pt\shortrightarrow$}
\newcommand{\shortarrow}[1]{%
  \mathrel{\text{\rotatebox[origin=c]{\numexpr#1*45}{\fixed@sra}}}
}
\makeatother

\begin{document}

\title{\LARGE{\bf An Observer-based Switching Algorithm for Safety \\\pno{under Sensor} \NotInt{Denial-of-Service} Attacks}}
\author{Santiago Jimenez Leudo, Kunal Garg, Ricardo G. Sanfelice and Alvaro A. Cardenas
\thanks{S. J. Leudo and R. G. Sanfelice are with the Department of Electrical and Computer Engineering, and  A. A. Cardenas is with the Department of Computer Science and Engineering,
University of California, Santa Cruz, CA 95064.
K. Garg is with the Department of Aeronautics and Astronautics, Massachusetts Institute of Technology, Cambridge, MA 02139.
      Email: {\tt\small \{sjimen28, ricardo, alacarde\}@ucsc.edu}, \tt\small\{kgarg\}@mit.edu}
\IfACC{ \thanks{ 
\pno{
Research partially supported by the NSF Grants no. ECS-1710621, CNS-2039054, and CNS-2111688, by the AFOSR Grants no. FA9550-19-1-0169, FA9550-20-1-0238, and FA9550-23-1-0145,
 by the AFRL Grant nos. FA8651-22-1-0017 and FA8651-23-1-0004,
by ARO Grant no. W911NF-20-1-0253, and by Fulbright Colombia - MinTIC.
 The views and conclusions of this document are those of the authors and should not be interpreted as representing the official policies of the ARO or the U.S. Government. The U.S. Government is authorized to reproduce and distribute reprints for Government purposes, notwithstanding any copyright notation herein.
 }
}}{}
}
\maketitle
\thispagestyle{empty}

\begin{abstract}
\pno{The design of} safe-critical control algorithms for systems under Denial-of-Service (DoS) \IfInt{and integrity}{} attacks on the system output \pno{is} studied in this work. We aim to address scenarios where attack-mitigation approaches are not feasible, and the system needs to maintain safety under adversarial attacks. We propose an attack-recovery strategy by designing a \pno{switching} observer and characterizing bounds in the error of a state estimation scheme by specifying tolerable limits on the time length of attacks. Then, we propose a \pno{switching} control algorithm that renders forward invariant a set for the \pno{observer}. Thus, by satisfying the error bounds of the state estimation, we guarantee that the safe set is rendered conditionally invariant with respect to a set of initial conditions. A numerical example \pno{illustrates} the efficacy of the approach.
\end{abstract}



\section{Introduction}




The security of Cyber-Physical Systems from a control-theoretic perspective is a growing area of research~\cite{chong2019tutorial}. Various types of attacks on a control system can occur, such as sensor data or system actuators getting compromised~\cite{teixeira2015secure,cardenas2008secure}.  Attackers can \IfInt{inject malicious (false) data in the output or actuator signals, thus compromising the integrity of the system~\cite{Mo2014IntegrityAttacks}. Alternatively, attackers can}{} disable the transmission of signals between devices, causing a Denial of Service (DoS) attack\IfACC{~\cite{Amin2009DoSNetControl}}{}. Such attacks can lead to violation of safety requirements, such as avoiding obstacles or keeping the system trajectories in a desired region of the state space\IfACC{~\cite{krotofil2014cps}}{}.
\NotACC{As an example, under a sensor-based DoS attack~\cite{Amin2009DoSNetControl}, Programmable Logic Controllers (PLCs) can compute control signals with \emph{stale data}~\cite{krotofil2014cps}, leading the system to unsafe regions.} 

Mitigating and responding to attacks is an active area of research. Some efforts have focused on developing robust \pno{observers} to prevent compromised data from affecting the \pno{feedback loops.} 
Secure estimation uses redundant observers to reconstruct the state, but they assume that only a certain number of sensors (in particular, less than half of the sensors) have been compromised~\cite{fawzi2014secure}. An alternative to reduce this level of redundancy \NotACC{necessary for guaranteeing a satisfactory reconstruction of the state }is to reject outliers with the use of robust statistics~\cite{leblanc2013resilient}. \IfACC{This approach}{When rejecting sensor signals, an alternative is to use \emph{virtual sensors} from digital twins of the system~\cite{piedrahita2017leveraging}. However, that} requires precise knowledge of the system's dynamical model\IfACC{.}{\> for provable guarantees for safety-critical systems.} The control \NotACC{or actuation} signal can also be constrained to prevent attackers from causing damages~\cite{kafash2018constraining}. However, such approaches \pno{may negatively} affect the system's performance. There is a plethora of work on resilient or robust control design, see, e.g., \cite{fawzi2014secure,yan2017resilient,bai2017kalman}, that focuses on system performance under attacks, however, without any consideration or guarantees on safety. 


Safety is perhaps the most important system property we need to maintain while undergoing an attack. Control barrier function (CBF)-based approaches can help design control algorithms for forward invariance of a safe set \cite{ames2014control}. The authors in \cite{clark2020control} introduce the notion of fault-tolerant CBF for handling attacks on stochastic systems. In \cite{amin2009safe}, the authors study safe control design under DoS attacks. 

In this paper\pno{,} we focus on the problem of safely recovering from output attacks, i.e., keeping the system trajectories in a safe set even under \NotInt{DoS} attacks\NotACC{\>on the system output}. 
The proposed formulation is applicable to several use cases with objectives including \pno{obstacle avoidance} and collision-free navigation for autonomous vehicles, reach-avoid control problems, surveillance, and convoy of multi-agent systems, among others. 
%
We \pno{propose a control scheme based on the information available, namely, the uncompromised outputs, that assures safety for systems with outputs experiencing DoS \IfInt{and integrity }{}attacks.} 
We consider scenarios {in which 
every attack has finite \pno{duration}, succeeded by an interval \pno{of time} without attacks.
We are interested in finding the set of initial conditions and the control action such that the state trajectory remains in the safe set at all times. \pno{During} attacks, the controller relies only on the uncompromised outputs, \pno{from which we generate an estimate of the state,} whereas the \pno{entire output} is used when attacks are not present.

\IfACC{
In this paper, we design a \pno{switching} observer scheme that uses the complete output information when there is no attack and uncompromised output information \pno{during} an attack in the sensors.
 We provide sufficient conditions \pno{involving key} properties of the system, such as the maximum tolerable length of the DoS attack and the minimum required length of the interval without an attack for recovery, \pno{guaranteeing that the} state estimation error remains uniformly bounded. 
Furthermore, we design CBF-based observer-based feedback laws to render a properly defined set forward invariant for the \pno{observer} so that with bounded estimation error, \pno{the system is safe. This is obtained provided conditional invariance of a set of interest with respect to a set of initial states.}
Due to space constraints, proofs and other details are not included and will be published elsewhere.
}{The main contributions of this paper are as follows.
\begin{itemize}
\item We design a \pno{switching} observer scheme that switches according to the presence of attack in the sensors. 
It uses the complete output information when there is no attack and uncompromised output information \pno{during} an attack.
\item We provide sufficient conditions \pno{involving key} properties of the system, such as the maximum tolerable length of the DoS attack and the minimum required length of the interval without an attack for recovery, \pno{guaranteeing that the} state estimation remains uniformly bounded.
\item We design CBF-based observer-based feedback laws to render a properly defined set forward invariant for the \pno{observer} so that with bounded estimation error, \pno{the system is safe. This is obtained provided conditional invariance of a set of interest with respect to a set of initial states.}
\end{itemize}}


\textbf{Notation.} {The symbols $\mathbb{R}$, $\mathbb{R}_{\geq 0}$, and $\mathbb{N}_{>0}$ denote the sets of real numbers, nonnegative reals, and positive natural numbers, respectively. Let $|x|$ be the \pno{E}uclidean norm of the vector $x$. 
Let $\overline{\mathcal{A}}$ denote the closure of the set ${\mathcal{A}}$.
Let $|A|$ be the \pno{induced matrix $2-$norm of $A$, $\textup{rank}(A)$ denote its rank,} and $\lambda_m(A), \lambda_M(A)$ denote the  eigenvalues with minimum and maximum real part, respectively. Let $\mathbb{B} \subset \mathbb{R}^n$ denote the \pno{closed} unit ball centered at the origin and $p+r\mathbb{B}$ the ball of radius $r\geq 0$ centered at $p\in \mathbb R^n$. We denote by $\tilde{\mathcal O} (C,A)$ the observability matrix of the pair $(C,A)$ and by $\tilde{\mathcal C} (A,B)$ the controllability matrix of the pair $(A,B)$.
}
\section{Preliminaries}
{
Consider the nonlinear system 
\begin{align}\label{eq: nl system}
\mathcal F : 
\quad\quad 
\dot z = F(t,z) \IfDist{+ d(t,x)},
\IfACC{\quad \quad}{\\}
y = H(t,z)
\end{align}
where $z\in 
\mathbb{R}^n$ is the system state, $y\in \mathbb R^p$ is the system output, 
 $F: \mathbb{R}_{\ge0} \times \mathbb{R}^{n} \rightarrow \mathbb{R}^n$ \pno{is the (potentially nonsmooth) flow map} and $H:\mathbb{R}_{\ge0} \times \mathbb{R}^{n} \rightarrow \mathbb{R}^p$ \pno{is the output map}.

 A solution to the system $\mathcal{F}$ is defined as follows.

{
\begin{Definition}[Solution to $\mathcal{F}$]
A \pno{locally absolutely continuous function $t \mapsto z(t)$} defines a solution to the 
system $\mathcal{F}$ in (\ref{eq: nl system}) 
  if
     $   \frac{d}{dt}z(t)= F(t,z(t))$
    \pno{for almost all $t \in \mathbb{R}_{\geq 0}$.}
\label{SolutiontocalH}
\end{Definition}}
  We say that a solution $z$ to $\mathcal{F}$ is maximal if it cannot be extended and we say it is complete when $\textup{dom} \> z = [0, \infty)$. 

\begin{Definition}[Safety] 
The system (\ref{eq: nl system}) is said to be safe with respect to $(X_0,X_u)$, with $X_0 \subset \mathbb{R}^{n} \setminus X_u$, if for each $z_0 \in X_0$, \pno{each} solution $t \mapsto z(t)$ to (\ref{eq: nl system}) \pno{with $z(0)=z_0$} satisfies $z(t) \in \mathbb{R}^{n}\setminus X_u$ for all $t \in \textup{dom}\> z$.
\end{Definition}

\begin{Definition}[Conditional invariance] 
A closed set $S \subset \mathbb{R}^{n}$ is said to be conditionally invariant for system (\ref{eq: nl system}) with respect to $M\subset S$ if, for each $z_0 \in M$, any solution $t \mapsto z(t)$ to (\ref{eq: nl system}) from $z_0$ satisfies $z(t) \in S$ for all $t \in \textup{dom}\> z$. 
\end{Definition}

It is immediate that the system (\ref{eq: nl system}) is safe with respect to $(X_0,X_u)$ if and only if the set $S := \mathbb{R}^n\setminus X_u$ is conditionally invariant for (\ref{eq: nl system}) with respect to  $X_0$.} \pno{For more details see \cite{198}.}

\section{Problem formulation}
\subsection{System Model}
Consider \pno{the} linear \pno{time-invariant} control system 
\begin{align}\label{eq: actual system}
\mathcal S : 
\quad\quad 
\dot x = Ax + Bu \IfDist{+ d(t,x)},
\IfACC{\quad \quad}{\\}
y = Cx
\end{align}
where $x\in \mathbb{R}^n$ 
is the system state, $y\in \mathbb R^p$ is the system output, 
$u\in \mathcal U$ is the control input, 
and $\mathcal U\subset \mathbb R^m$. Here, \pno{$A\in \mathbb R^{n\times n}$, $B\in \mathbb R^{n\times m}$, 
\NotDist{and} $C\in \mathbb R^{p\times n}.$}

\subsection{Attack Model}
In this work, we consider attacks on the system output $y$. In particular, we consider an attack where a subset of the components of the system output is compromised. Under such an attack model, the \textit{measured} system output {$\bar y$} takes the form
%
\begin{equation}\label{eq: attack model}
    \bar {y} = 
    (y_s , y_a) 
\end{equation}
where 
 $  y_s=\tilde Cx,
$ and, \pno{for each solution $t \mapsto x(t)$ to \eqref{eq: actual system},} 

\begin{align}\pno{
    y_a(t) = \begin{cases}\bar C x(t) & \textrm{if} \; t\notin \mathcal T_a, \\ Y(t,x(t)) & \textrm{if} \; t\in \mathcal T_a\end{cases}}
\end{align}
The quantity $\tilde Cx$ denotes the \textit{secured} output components that \textit{cannot} be attacked with $\tilde C\in \mathbb R^{\tilde p\times n}$ and $0\leq \tilde p<p$, $\bar Cx$ denotes the \textit{vulnerable} output components that \textit{can} be attacked with $\bar C\in \mathbb R^{(p-\tilde p)\times n}$ such that $C = \begin{bmatrix}\tilde C\\ \bar C\end{bmatrix}$, and $Y:\mathbb R_{\geq 0}\times\mathbb R^n\rightarrow\mathbb R^{p-\tilde p}$ denotes the attacked output signal.
We denote with $\mathcal T_a\subset\mathbb R_{\geq 0}$ the set of times when an attack is present on the system output, \pno{which is assumed to be known provided a DoS attack detection mechanism}. 
The attack model \eqref{eq: attack model} captures \IfInt{both}{} Denial-of-Service (DoS) \IfInt{and integrity}{} attacks on the system output\IfInt{, and thus, models a large class of cyber attacks \cite{Amin2009DoSNetControl,Mo2014IntegrityAttacks}}{}. Let $[t_1^i, t_2^i)$ with $t_2^i>t_1^i\geq 0$ denote the interval of time \pno{over which} the \pno{$i-$th} DoS attack \pno{occurs}, with \pn{$i \in \mathbb{N}_{>0}$.} 
\pno{Define $\mathcal T_a \coloneqq \bigcup\limits_i[t_1^i, t_2^i)$,
$\mathcal T_1 = \bigcup\limits_i\{t_1^i\}$, and $\mathcal T_2 = \bigcup\limits_i \{t_2^i\}$ as the intervals of attack, and the sets of the starting and ending time instants of attacks, respectively.
To provide sufficient conditions to guarantee safety, we characterize the attacks by defining}
\IfInt
{\begin{subequations}
\begin{align}
    T_a & \coloneqq \max_{i\in \{1,2,\dots\} }(t_2^i-t_1^i),\\
    T_{na} & \coloneqq \min_{i\in \{2,3,\dots\}}(t_1^i-t^{i-1}_2),
\end{align}
\end{subequations}}{
 $T_a  \coloneqq \max_{i\in \{1,2,\dots\} }(t_2^i-t_1^i)$ and $
    T_{na}  \coloneqq \min_{i\in \{2,3,\dots\}}(t_1^i-t^{i-1}_2)$
}
as the maximum length of the DoS attack and the minimum length of the interval without an attack, 
respectively. Notice that $t_2^0:=0$, and when $t_1^1>0$, we have $t_1^1\geq T_{na}$.

\subsection{Problem Statement}
Given a nonempty, closed set $S\subset\mathbb R^n$, referred to as \pno{ the {\em safe}} set, the problem to solve is the design of an algorithm such that the set $S$ is conditionally invariant for (\ref{eq: actual system}) with respect to the set $X_0$.
\IfDist{We make the following assumption on the unmodeled dynamics $d$ in \eqref{eq: actual system}:
\begin{Assumption}\label{assum: d bound}
There exists $\gamma>0$ such that $|d(t, x)|\leq \gamma$ for all $t\geq 0$ and $x\in \mathcal D$.
\end{Assumption}}
%
%
Formally, the control design problem studied in this paper is stated as follows.

\begin{Problem}\label{Problem 1}
Given system \eqref{eq: actual system}\IfDist{ with unmodeled dynamics $d$ that satisfies Assumption \ref{assum: d bound}}, 
a closed set $S \subset \mathbb{R}^n$, and the attack model in \eqref{eq: attack model}, 
\begin{enumerate}
    \item Find a set of initial states $X_0\subset S$, 
    and
    \item Design a control 
    law {$\kappa$} assigning the input $u$ of \eqref{eq: actual system} \pno{using measurements of $\bar y$}
\end{enumerate}
    such that\pno{,} for each $x_0 \in X_0$, 
    the 
    solution to the resulting closed-loop system, 
    namely $t \mapsto x(t)$, with $x(0)=x_0$, 
    satisfies $x(t) \in S$ 
    for all $t\geq 0$.
\end{Problem}


\subsection{Proposed Solution}

To solve Problem $(\star)$, we propose the design of an observer-based feedback law that induces conditional invariance of $S$ with respect to $X_0$. 
Most CBF-based methods for forward invariance rely on measurement of the entire state \cite{ames2017control}. 
We propose to employ a state estimator that reconstructs the system state using the measured output $\bar y$. The observer is given as
\begin{align}\label{eq: pred model}
    \dot{\hat x} \NotACC{&}
    = A\hat x + Bu + g(\bar y,\hat y), \IfACC{\quad \quad}{\\}
    \hat y \NotACC{&}
    = C\hat x,
\end{align}
\pno{where $\hat x\in \mathbb R^n$ is the estimate of $x$} and $g:\mathbb R^p\times\mathbb R^p\rightarrow\mathbb R^n$ is the innovation term to be designed \pno{such that $g(\bar{y},\hat{y})=0$ at $\bar y = \hat y$}. 
When the system output is under an attack according to the attack model \eqref{eq: attack model}, the actual output information is not available to the state observer. Thus, the observer needs to take into account the attacks on the system output. To this end, we design an observer that uses the \textit{complete output} vector when there is no attack and only the \textit{non-attacked output} components when the system output is under attack. More specifically, the proposed observer under the attack model \eqref{eq: attack model} is given as
\begin{align}\label{eq: hat x switch general}
    \dot {\hat x} = \begin{cases} A\hat x + Bu + g_1(Cx,C\hat x) & \textrm{if} \quad  t\notin \mathcal T_a, \\
    A\hat x + Bu + g_2(\tilde Cx, C\hat x) & \textrm{if} \quad t\in \mathcal T_a \end{cases}
\end{align}
where $g_1, g_2:\mathbb R^p\times \mathbb R^p\rightarrow\mathbb R^n$ are to be designed. 
Given a set $\mathcal{T}_a \subset \mathbb{R}$, the 
feedback law $\kappa$ assigning $u$ is defined as 
\begin{align}\label{eq: u switch} 
    \kappa(t,\hat x, y) = \begin{cases}
    \kappa_1(\hat x, y) & \textrm{if} \quad t\notin \mathcal T_a, \\
    \kappa_2(\hat x, y) & \textrm{if} \quad t\in \mathcal T_a,
    \end{cases}
\end{align}\\
where $\kappa_1, \kappa_2: \mathbb R^{n} \times \mathbb{R}^p \rightarrow\mathbb R^m$ are functions to be designed under \textit{nominal} operation (i.e., when the system is not under an attack) and \pno{under attack,} respectively.
Notice that the closed-loop system
resulting from the composition of \eqref{eq: actual system} and \eqref{eq: hat x switch general} with $\kappa$ as in \eqref{eq: u switch} can be expressed \pno{as in} \eqref{eq: nl system} with $z = (x, \hat x)$.
%

We make the following assumption on $\mathcal{S}$ in \eqref{eq: actual system}.
\begin{Assumption}\label{assum A B cont}
The pair $(A,B)$ is controllable and the pair $(C,A)$ is {detectable.}
\end{Assumption}

Based on the structure of the observer in \eqref{eq: actual system} and the {observer-based feedback law} in \eqref{eq: u switch}, the approach followed in this paper for \pno{safety under attacks for} system \eqref{eq: actual system} is as follows.\\
\textbf{Approach}: 
Given a closed set $S \subset \mathbb{R}^n$, the system \eqref{eq: actual system}
,\IfDist{ with unmodeled dynamics $d$ satisfying Assumption \ref{assum: d bound},} and the attack model \eqref{eq: attack model}, 
our approach is to compute sets $X_0, \hat X_0,\hat S_0\subset S$ and design functions $g_1,g_2$ for the observer in \eqref{eq: hat x switch general} and functions $\kappa_1, \kappa_2$ for the {observer-based feedback law} $\kappa$ as in \eqref{eq: u switch} such that
\pno{each} solution pair $t \mapsto (x(t), \hat x(t))$ to the closed-loop system resulting from the composition of \eqref{eq: actual system} and \eqref{eq: hat x switch general} with  $\kappa$ satisfies \pno{the following properties:}
\begin{itemize}
    \item[1)] For each $t_0\in \mathcal T_1$ such that $x(t_0)\in X_0$ and $\hat x(t_0)\in \hat X_0$, the $x$ component of the resulting closed-loop solution satisfies $x(t)\in S$ for all $t\in [t_0, t_0+T_a)$;
    \item[2)] For each $t_0\in \mathcal T_2$ such that $x(t_0)\in S$ and $\hat x(t_0)\in  \hat S_0$, 
    and for $\hat t_0 =\max \{t_0, \inf_{t\geq t_0} \mathcal T_1 \}$,
    the $x$ component of the resulting closed-loop solution satisfies 
    $x(\hat t_0)\in X_0$ and $x(t)\in S$ for all $t\in [t_0, \hat t_0)$.
   
%
%
\end{itemize}
\begin{Remark}
\pno{The sets $\hat{X}_0$ and $\hat{S}_0$ denote the sets of estimates before and after an attack, respectively. We will design these sets in the next section.}
Item \pno{1} in our solution approach encodes conditional invariance of the set $S$ 
for system \eqref{eq: actual system} with respect to $X_0$, under an attack with maximum duration.
 Upon the requirement of the state to be in $S$ at the end of every attack, item \pno{2} encodes safety of  system \eqref{eq: actual system} with respect to $(X_0, \mathbb{R}^n \setminus S)$ during the 
 time-intervals with no attacks,
 {and the state to be in $X_0$ 
 at the beginning of the next attack.}
%
\end{Remark}

\NotACC{In the next section, we present the design of the observer in \eqref{eq: hat x switch general}, and in the following section, we present the design of the {observer-based feedback law} in \eqref{eq: u switch}.}

\section{Switching Observer Design}
Under an attack on the system output of the form \eqref{eq: attack model}, it might not be possible to reconstruct the state \pno{of \eqref{eq: actual system}} for a full-state feedback control design. Specifically, under the considered attack model, the rank of the observability matrix $\tilde{\mathcal O}$ for the pair $(\tilde C,A)$, namely, $\textup{rank}(\tilde{\mathcal O}) = \tilde n$, potentially \pno{smaller than $n$.} Thus, there might be \pno{$n-\tilde n>0$} eigenvalues in the closed right-half plane for the \pno{dynamics of the estimation error resulting for} any observer design under attack. Keeping this in mind, \pno{the \pno{switching} observer in \eqref{eq: hat x switch general} is defined as}
\begin{align}\label{eq: hat x switch}
    \dot {\hat x} = \begin{cases} A\hat x+ Bu  + L(Cx-C\hat x) & \textrm{if} \quad  t\notin \mathcal T_a, \\
    A\hat x + Bu +\tilde L(\tilde Cx-\tilde C\hat x)& \textrm{if} \quad t\in \mathcal T_a,\end{cases}
\end{align}
where \pno{$L\in \mathbb{R}^{n\times p}$ and} $\tilde L \in \mathbb R^{n\times \tilde p}$ is such that $\tilde n$ (with $\tilde n\leq n$) eigenvalues of the matrix $A-\tilde L\tilde C$ lie in the open left-half plane. 
On the other hand, since $(C, A)$ is detectable under Assumption \ref{assum A B cont}, we can design $L$ such that all the eigenvalues of $(A-LC)$ are in the open left-half plane.
Now, define $e = x -\hat x$ as the \pno{estimation} error to obtain the error dynamics given as  
\begin{align} \label{eq:errordyn}
    \dot e = \begin{cases} (A-LC)e \IfDist{+d(t,x)} & \textrm{if} \quad t\notin \mathcal T_a, \\
    (A-\tilde L\tilde C)e\IfDist{+d(t,x)} & \textrm{if} \quad t\in \mathcal T_a\end{cases}
\end{align}
with $ e(0) = x(0) - \hat x(0)$. Next, we analyze the error bounds when there is no attack, i.e., at each $t\notin \mathcal T_a$.

\subsection{Analysis under No Attacks}

Consider the starting instant of an interval during which there is no attack on the system output, namely 
$t_2^i \in
\mathcal{T}_2\cup \{0\}$, with $i \in \mathbb{N}$
. 
{The following result is the initial step to guarantee conditional invariance of $S$ with respect to $X_0$ 
for the  system \eqref{eq: actual system} when there are no attacks.}

\begin{Lemma}
\pno{Given system (\ref{eq: actual system}), suppose Assumption \IfDist{\ref{assum: d bound}-}\NotDist{\ref{assum A B cont}} 
holds.}
For given $\IfDist{\gamma,}
T_{na}, \bar e_0>0$,
an associated \pno{observer} (\ref{eq: hat x switch}),
and corresponding error dynamics (\ref{eq:errordyn}),
%
\pno{if at the} $i-$th interval of no attacks with {$i \in \mathbb{N}$,} 
%
{ $|e(t_2^i)| \leq \bar e_0$ with $t_2^i\in \mathcal{T}_2$,  then}
the \pno{state estimation} error satisfies 
$|e(t)|\leq \pn{\gamma_1}(t{-t_2^i})
\bar e_0$ for all $t \in  [t_2^i, t_1^{i+1}]$, 
where  
\begin{align}\label{eq: C1}
    \pn{\gamma_1}(t)
    \coloneqq c_1 \exp\left({-\bar \lambda_1
    t}\right)
\end{align}
with 
$\bar \lambda_1 = \frac{\lambda_m(Q)}{2\lambda_M(P)}$, ${c_1} = \sqrt{\frac{\lambda_M(P)}{\lambda_m(P)}}$, 
\pno{ and $L$ such that}
for some \pno{symmetric} positive definite matrices $P$ \pno{and} $Q$, { $-Q=(A-LC)^\top P + P(A-LC)$} \pno{holds}.
 \label{Lemma: Bound un attack}
\end{Lemma}

\NotACC{
\begin{proof}
Since the pair $(C, A)$ is detectable, it follows that there exist positive definite matrices $P, Q$ such that $V:\mathbb R^n\rightarrow\mathbb R_{\geq 0}$, defined as $V(e) \coloneqq e^TPe$, satisfies
$$
\begin{aligned}
\dot{V}(e)&=\dot{e}^{\top} P e+e^{\top} P \dot{e}\\ &=[(A-L C) e\IfDist{+d}]^{\top} P e+e^{\top} P[(A-L C) e \IfDist{+d}] \\
&=e^{\top}(A-L C)^{\top} P e\IfDist{+d^{\top} P e}+e^{\top} P(A-L C) e\IfDist{+e^{\top} P d} \\
&=-e^{\top} Q e\IfDist{+2 e^{\top} P d}.
\end{aligned}
$$
%
\IfDist{Under Assumption \ref{assum: d bound}, and g}\NotDist{G}iven that $ \lambda_m(Q) e^\top  e  \leq e^\top Q e$ for all $e \in \mathbb{R}^n$, we obtain
\begin{align}
    \dot V(e)\leq -|e|^2\lambda_m(Q) \IfDist{+ 2|e||P| \pn{\gamma}}.
    \label{bound1Vd}
\end{align}
Since $P$ is symmetric, we have 
\begin{equation} 
\lambda_m(P)|e|^2 \leq V(e) \leq \lambda_M(P)|e|^2,
\label{eq:boundP}
\end{equation}
for all $e\in \mathbb R^n$, which, together with \eqref{bound1Vd} implies that 
$\dot{V}(e)\leq -\frac{\lambda_m(Q)}{\lambda_M(P)}V(e)    \IfDist{\quad \quad e(t)\in  E}$. Let $t \mapsto e(t)$ be a solution of \eqref{eq:errordyn} and consider the interval $[t_2^i, t_1^{i+1}]$ for $i \in \mathbb{N}_{>0}$. It follows that 
$V(e(t))\leq V(e(t_2^i)) \exp\left(-\frac{\lambda_m(Q)}{\lambda_M(P)}(t-t_2^i) 
\right)     \IfDist{\quad \quad e(t)\in  E}$ for all  $t \in [t_2^i, t_1^{i+1}]$.
Since $V(e(t_2^i)) \leq \lambda_M (P) |e(t_2^i)|^2$, using the left inequality in (\ref{eq:boundP}), it follows that
\begin{align*}\lambda_m (P) |e(t)|^2\leq \lambda_M (P) \exp\left({-\frac{\lambda_m(Q)}{\lambda_M(P)} (t{-t_2^i})
}\right) |e{(t_2^i)}|^2    \IfDist{\\ \quad \quad e(t)\in  E}\end{align*}
{for all $t \in [t_2^i, t_1^{i+1}]$}, 
which thanks to $|e{(t_2^i)}|\leq \bar e_0$, 
implies that
\begin{align*}
    |e(t)|\leq c_1 \exp \left({-\bar \lambda_1(t{-t_2^i})
    }\right)|e{(t_2^i)}|
    \IfDist{\quad \quad e(t)\in  E}
    \leq \gamma_1(t{-t_2^i})\bar e_0,
\end{align*}
for all  $t \in [t_2^i, t_1^{i+1}]$,
completing the proof. 
\end{proof}
}

\IfDist{Furthermore, given that the rate of growth of the bound is negative, it holds that for each $e(t_0)\in \mathbb R^n\setminus E$, $e(t)\in \mathbb R^n\setminus E$ for all $t\geq t_0$ without an attack. }

Notice that the above analysis (with a nominal Luenberger observer) can be used to show that starting from $e(t_2^i) $ {with $t_2^i \in \mathcal{T}_2 \cup \{0\}, i\in \mathbb{N}$}, the error exponentially converges to  $\delta \mathbb B$ in time $T_{na}$, where
$
    \delta = 
    \gamma_1(T_{na})|e(t_2^i)|
$, and stays in that ball \pno{until} the next attack starts at $ t^{i+1}_1$. 


\begin{Remark}
The Luenberger observer used when there are no attacks is just one choice of a \pno{state estimator}. It is \pno{also} possible to use a finite-time stable \pno{state estimator} \cite{shen2008semi}, or any other observer 
that has faster convergence guarantees. 
\end{Remark}

\subsection{Analysis under Attacks}
During the attack on the output, we use a different observer gain designed for the pair \pno{$(\tilde C, A)$}. Since it might not be possible to place all the eigenvalues of $A-\tilde L\tilde C$ in the open left-half plane, {the matrix $\tilde L$ in \eqref{eq: hat x switch} can be designed to minimize the maximum eigenvalue of $A-\tilde L\tilde C$, which minimizes the rate of growth of the error during \pno{attacks}. 
Based on $\tilde L$, we compute the maximum growth rate possible in the estimation error $e$ during intervals of attacks in the system output, 
assuming a worst-case attack.

Under the attack model \eqref{eq: attack model}, a subset of the state space may still be \pno{detectable} for the pair $(\tilde C, A)$. Thus, under the \pno{observer} \eqref{eq: hat x switch} for $t\in \mathcal T_a$, it is possible that some of the eigenvalues of the matrix $A-\tilde L\tilde C$ are in the open left-half plane. \pno{To bound} the error growth during the attack, we \pno{consider the general case in which we can decompose} the matrix $A-\tilde L\tilde C$ into submatrices $\hat A_{11}$ and $\hat A_{22}$, such that the eigenvalues of $\hat A_{11}$ are in the open left-half plane. To this end, let $\Phi \in \mathbb{R}^{n \times n}$ be an invertible matrix consisting of the generalized eigenvectors of the matrix $A-\tilde L\tilde C$ such that 
\small{
\begin{align}\label{eq: Phi mat}{\small
\Phi^{-1} (A-\tilde L\tilde C)\Phi = \begin{bmatrix}\hat A_{11} & 0_{\tilde n \times (n-\tilde n)} \\ 0_{(n-\tilde n) \times \tilde n} & \hat A_{22}\end{bmatrix}}
\end{align}}\normalsize
where $\hat A_{11}$ and $\hat A_{22}$ are Jordan blocks such that $\lambda_M (\hat A_{11})<0$ and $0_{p\times q}\in \mathbb R^{p\times q}$ is a matrix consisting of zeros\footnote{Note that it is always possible to find the Jordan form of the matrix $A-\tilde L\tilde C$, even when it is not diagonalizable.}. Also, let {$\Phi^{-1}=\big [\hat{\Phi}_1^\top,  \hat{\Phi}_2^\top \big ]$}, and define the change of coordinates $z=\Phi^{-1} e$. Then, $e=\Phi z$, and in the new coordinates, the error dynamics are expressed as
\pno{\small
\begin{align*}
    \dot z = \Phi^{-1} \dot e 
        & =  \begin{bmatrix}\hat A_{11} & 0_{\tilde n \times (n-\tilde n)} \\ 0_{(n-\tilde n) \times \tilde n} & \hat A_{22}\end{bmatrix} z \IfDist{+ \Phi^{-1} d(t,x)}.
\end{align*}}

Define $z=(z_{11},z_{22})$, where $z_{11} \in \mathbb{R}^{\tilde n}$ and $z_{22} \in \mathbb{R}^{n-\tilde n}$ so that we have
\IfACC
{
$\dot{z}=(\dot z_{11},\dot z_{22})  = (\hat A_{11} z_{11}, \hat A_{22} z_{22}).$
}
{
\begin{align}
 \dot z_{11} &= \hat A_{11} z_{11} \IfDist{+ \Phi_1 d(t,x)} \label{eq:dz11}  \\
      \dot z_{22} &= \hat A_{22} z_{22}. \IfDist{+ \Phi_2 d(t,x)} \label{eq:dz22}
\end{align}
}
We can now state the following result \pno{providing a} bound on the state estimation error under attacks. 
\begin{Lemma}
\pno{Given system (\ref{eq: actual system}), suppose Assumption \IfDist{\ref{assum: d bound}-}\NotDist{\ref{assum A B cont}} 
holds.}
For given $
T_{a}, \bar{e}_0>0$, 
\pno{ an associated observer} (\ref{eq: hat x switch}), 
and corresponding error dynamics (\ref{eq:errordyn}),
%
\pno{if at the} $i-$th interval of attack with {$i \in \mathbb{N}_{>0}$} 
and maximum length $T_{a}$,
{ $|e(t_1^i)| \leq \bar e_0$ with $t_1^i\in \mathcal{T}_1$,  then}
%
%
%
the state estimation error satisfies $|e(t)|\leq { 
\gamma_2(T_a) \bar e_0}$ for all $t\in [t_1^i, t_2^i]$, 
where 
\begin{align}\label{eq: C2}
    \gamma_2(T_a) \coloneqq \max\limits_{t\in [0, T_a]} \hat c_1 \exp\left({-\hat\lambda_1 t}\right) + \hat c_2 \exp\left({\hat\lambda_2 t}\right)
\end{align}
with 
\footnotesize{
\IfInt{
\begin{align}\label{eq: c1 c2 params}
    & \hat c_1 = |\Phi||\hat \Phi_1|\sqrt{\frac{\lambda_M(\hat P)}{\lambda_m(\hat P)}}, &
    \hat c_2 & = |\Phi||\hat \Phi_2|,
    \\ 
    & \hat\lambda_1 = \frac{\lambda_m(\hat Q)}{2\lambda_M(\hat P)},
    &\hat \lambda_2 & = |\hat A_{22}|,
\end{align}
}{
$$
\hat c_1 = |\Phi||\hat \Phi_1|\sqrt{\frac{\lambda_M(\hat P)}{\lambda_m(\hat P)}}, 
\>\>\>\>
    \hat c_2  = |\Phi||\hat \Phi_2|,
    \>\>\>\>
     \hat\lambda_1 = \frac{\lambda_m(\hat Q)}{2\lambda_M(\hat P)},
     \>\>\>\>
    \hat \lambda_2  = |\hat A_{22}|,
$$}}\normalsize
\noindent \pno{and $\tilde L$ such that} for some \pno{symmetric} positive definite matrices $\hat P$ \pno{and} $ \hat Q$, $-\hat Q=\hat{A}_{11}^{\top} \hat P+\hat P\hat{A}_{11}$ \pno{holds}.
%
%
 \label{Lemma: Bound u attack}
\end{Lemma}
\NotACC{
\begin{proof} 
Consider the dynamics \eqref{eq:dz11}. Since by construction, the eigenvalues of the matrix $\hat{A}_{11}$ are in the open left-half plane, it follows that there exist positive definite matrices  $\hat P, \hat Q$ such that $\hat V:\mathbb R^l\rightarrow\mathbb R_{\geq 0}$ defined as $\hat V(z_{11}) \coloneqq z_{11}^\top \hat P z_{11}$ satisfies
\begin{align*}
    \dot {\hat V}(z_{11})  = -z_{11}^\top \hat Q z_{11} \IfDist{+ 2 z_{11}^\top \hat P \Phi_1 d(t,x)}
\end{align*}
%
\IfDist{Under Assumption \ref{assum: d bound}, and g}\NotDist{G}iven that $ \lambda_m(\hat Q) z_{11}^\top  z_{11} \leq z_{11}^\top \hat Q z_{11} $ for all $z_{11} \in \mathbb{R}^{\tilde n}$, we obtain
\begin{align}
    \dot{ \hat V}(z_{11})\leq -|z_{11}|^2\lambda_m(\hat Q) \IfDist{+ 2|z_{11}||\hat P| |\Phi_1| \gamma}.
    \label{bound1Vhd}
\end{align}
Let $t\mapsto (z_{11}(t), z_{22}(t))$ denote a solution of \eqref{eq:dz11}-\eqref{eq:dz22}. 
Denote 
$\hat \lambda_1 = \frac{\lambda_m(\hat Q)}{2\lambda_M(\hat P)}$ and 
$\breve c_1=\sqrt{\frac{\lambda_M(\hat P)}{\lambda_m(\hat P)}}$ to obtain that
\begin{align*}
    |z_{11}(t)|\leq \breve c_1 \exp \left({-\hat \lambda_1(t-t_1^i)}\right)|z_{11}(t_1^i)|
    \IfDist{\quad\\ z_{11}(t)\in  \hat E}.
\end{align*}
On the other hand, for \eqref{eq:dz22},
%
given that $|\hat A_{22} z_{22} \IfDist{+ \Phi_2 d(t,x)}| \leq  |\hat A_{22}| |z_{22}| \IfDist{+ |\Phi_2| \gamma }$, we have
\begin{align*}
    |z_{22}(t)|\leq {|z_{22}(t_1^i)|}\IfDist{+\frac{|\Phi_2| \gamma} }{}\exp(|\hat{A}_{22}| (t-t_1^i)) \IfDist{- \frac{|\Phi_2| \gamma }{|\hat{A}_{22}|}}
\end{align*}
\IfDist{So 
\begin{align*}
    |z_{22}(t)|\leq \exp(|\hat{A}_{22}| (t-t_0)) \left({|z_{22}(t_0)|}+\frac{|\Phi_2| \gamma }{|\hat{A}_{22}|}\right)
\end{align*}}
Thus, by denoting ${\hat c_1=|\Phi||\hat \Phi_1|\breve c_1},
$ and 
${\hat c_2=|\Phi||\hat \Phi_2|}, \hat\lambda_2=|\hat{A}_{22}| $, in the original coordinates we have
   \begin{align*}
    |e(t)| \leq& |\Phi| (|z_{11}(t)| + |z_{22}(t)|)\\
    \leq&  {|\Phi|\breve c_1
     |z_{11}(t_1^i)|
     \exp \left({-\hat\lambda_1(t-t_1^i)}\right)  }
     \\
    &+{|\Phi|
    {|z_{22}(t_1^i)|}\IfDist{+\frac{|\hat \Phi_2| \gamma} }{}}
    \exp(|\hat{A}_{22}| (t-t_1^i)) 
    \\ 
    =
    &  {|\Phi|\breve c_1
    |{\hat{\Phi}_{1}}e(t_1^i)|
    \exp \left({-\hat\lambda_1(t-t_1^i)}\right)  }
    \\
    &+{|\Phi|
    {|{\hat{\Phi}_2}e(t_1^i)|}\IfDist{+\frac{|\Phi_2| \gamma} }{}}
    \exp(\hat\lambda_2 (t-t_1^i)) 
    \\
    \leq & \left({\hat c_1} \exp ({-\hat\lambda_1(t-t_1^i)})  +\hat c_2 \exp(\hat\lambda_2 (t-t_1^i)) \right)|e(t_1^i)|
       \IfDist{
       \\ 
    &+\frac{\hat c_2 \gamma }{\lambda_2} \exp(\lambda_2 (t-t_1^i))}
 \end{align*}
Thus, thanks to $|e(t_1^i)| \leq \bar e_0$, for all $t\in [t_1^i, t_2^i]$ it follows that
\begin{align*}
    |e(t)|\leq 
     \max_{t\in [0, T_a]}\left({\hat c_1} \exp ({-\hat\lambda_1 t}) 
     +\hat c_2 \exp(\hat\lambda_2 t) \right)\bar e_0
\end{align*}
which completes the proof. 
\end{proof}
}
\NotACC{Note that it is possible that $\tilde n = 0$, i.e., all the eigenvalues of the matrix $A-\tilde L\tilde C$ are in the closed right-half plane. In that case, $\hat c_1 = 0$ in \eqref{eq: C2}.} 
\pno{ \subsection{Global Bound on Estimation Error}
Before we state the first main result of the paper, we make the following assumption on the initial state estimation error. 
\begin{Assumption}
The closed set $S\subset \mathbb{R}^n$ 
is such that there exists $\bar E>0$ {such that}, for the initial state $x(0)\in S$ 
and initial estimate 
$\hat x(0) \in S$, 
the error satisfies 
$|e(0)|=|x(0)- \hat x(0)|\leq \bar E$.
\label{ass:bounderror}
\end{Assumption}
A pre-defined initial error bound helps us guarantee the existence of a \pno{switching} observer of the form \eqref{eq: hat x switch general} such that \pno{safety is guaranteed}.

Now, we provide a result on bounds on the state estimation error under the proposed switching observer algorithm.
}

\begin{Theorem}
\pno{
Given system (\ref{eq: actual system}), suppose Assumptions \IfDist{\ref{assum: d bound}-}\NotDist{\ref{assum A B cont}} and \ref{ass:bounderror} hold for $\bar{E}>0$.
}
For given $T_{na}, T_a>0$,
an associated observer (\ref{eq: hat x switch}),
and corresponding error dynamics (\ref{eq:errordyn}), let $c_1,\bar \lambda_1,\hat c_1, \hat c_2, \hat\lambda_1, \hat\lambda_2>0$ be defined as per Lemma \ref{Lemma: Bound un attack} and Lemma \ref{Lemma: Bound u attack}.
    If $T_{na}$ and $T_a$ are such that
$\gamma_1(T_{na})\gamma_2(T_a) \leq 1$ with $\gamma_1$ as in \eqref{eq: C1} and $\gamma_2$ as in \eqref{eq: C2}, then 
    \pn{$|e(t)|\leq  \gamma_1(0)\gamma_2(T_a) \bar E$} for all $t\geq 0$. In addition,
    \begin{itemize}
    \item\pno{if there is an attack at time $t=0$, then
    $|e(t)|\leq  \bar E$ for all $t\in \mathcal{T}_1 \cup \{0\}$, and
    \item if the first attack is launched after at least $T_{na}$ seconds, then
    $|e(t)|\leq  \bar E$ for all $t\in \mathcal{T}_2 \cup \{0\}$.}
    \end{itemize}
\label{Th:TBoundError}
\end{Theorem}
\NotACC{
\begin{proof}
From Lemma \ref{Lemma: Bound un attack}, we have that for each 
interval without attacks starting at $t_2^i \in \mathcal{T}_2$ and ending at $t_1^{i+1} \in \mathcal{T}_1$, with $i \in \mathbb{N}$,
the estimation error satisfies 
\begin{multline}
|e(t_1^{i+1})|\leq c_1 \exp\left(-\bar \lambda_1 (t_1^{i+1} - t_2^i)
\right) |e(t_2^i)| \IfDist{+ \tilde c_2\gamma \exp(\lambda_2 T_a)\frac{1}{\lambda_2}}
\\
\leq c_1 \exp\left(-\bar \lambda_1 T_{na}\right) |e(t_2^i)|
=
\gamma_1(T_{na})|e(t_2^i)|
\label{eq: proofth1}
\end{multline}
The right-hand bound holds for any $t_1^{i+1}\geq T_{na}$ thanks to $c_1,\bar \lambda_1 >0$ 
in Lemma \ref{Lemma: Bound un attack}, namely, thanks to the exponential being decrescent.
%
For each attack interval starting at $t_1^{i} \in \mathcal{T}_1$ and ending at $t_2^{i} \in \mathcal{T}_2$, with $i \in \mathbb{N}_{>0}$,
the estimation error satisfies 
\begin{multline}
{\scriptstyle 
|e(t_2^{i})|\leq \max\limits_{t\in [t_1^i, t_1^i+T_a]} \left\{{\hat c_1} \exp ({-\hat\lambda_1(t-t_1^i)})  +\hat c_2 \exp(\hat\lambda_2 (t-t_1^i)) \right\}|e(t_1^i)|}
\\
{\scriptstyle
=
\max\limits_{t\in [0, T_a]} \left\{{\hat c_1} \exp (-\hat\lambda_1(t))  +\hat c_2 \exp(\hat\lambda_2 (t)) \right\}|e(t_1^i)|
=\gamma_2(T_a)|e(t_1^i)|}
\label{eq: proofth2}
\end{multline}
Thus, 
if there is an attack at the initial time, namely $t_1^1=0$,
from Assumption \ref{ass:bounderror}, we have $|e(t_1^1)|\leq \bar E$, 
from \eqref{eq: proofth2} we have $|e(t_2^1)|\leq \gamma_2(T_a) \bar E$, and 
from  
\eqref{eq: proofth1} we have $|e(t_1^2)|\leq \gamma_1(T_{na})\gamma_2(T_a) \bar E$. {If
$T_{na}$ and $T_a$ are such that $\gamma_1(T_{na})\gamma_2(T_a) \leq 1$, then $|e(t_1^2)|\leq \bar E$,} namely, at the beginning of the second attack, the error will be bounded by $\bar E$.
Recursively, this implies that $|e(t)| \leq \bar E$ at the beginning of every attack, namely, for all $t\in \mathcal{T}_1 \cup \{0\}$.
Notice that 
the error is bounded during the first interval of attack by 
$|e(t)| \leq \gamma_2(T_a)\bar E$ for all $t \in [0, t_2^1]$.
Given that 
$\max\limits_{t\in [0, T_{na}]} \gamma_1(t)=\gamma_1(0)$,  we have that
$|e(t)|\leq \gamma_1(0)\gamma_2(T_a)\bar E$ for all $t\in [0,t_1^2]$.
Recursively, 
this implies that $|e(t)| \leq \gamma_1(0)\gamma_2(T_a)\bar E$ for all $t \geq 0$.

On the other hand, if there is no attack at the beginning, we have at least $T_{na}$ seconds before the first attack is launched, namely $t_1^1>T_{na}$, 
and from Assumption \ref{ass:bounderror}, we have $|e(t_2^0)|\leq \bar E$, 
from \eqref{eq: proofth1} we have $|e(t_1^1)|\leq \gamma_1(T_{na}) \bar E$, and 
from \eqref{eq: proofth2}, $|e(t_2^1)|\leq \gamma_1(T_{na})\gamma_2(T_a) \bar E$. If
$T_{na}$ and $T_a$ are such that $\gamma_1(T_{na})\gamma_2(T_a) \leq 1$, then $|e(t_2^1)|\leq \bar E$, i.e., at the end of the first attack, the error will be bounded by $\bar E$. 
Recursively, this implies that $|e(t)| \leq \bar E$ at the end of every attack, namely, for all $t\in \mathcal{T}_2 \cup \{0\}$.
Notice that since 
$\max\limits_{t\in [0, T_{na}]} \gamma_1(t)=\gamma_1(0)$,  the error is bounded during the first interval without attacks by 
$|e(t)|\leq \gamma_1(0)\bar E$ for all $t\in [0,t_1^1]$. 
Given that the function $\hat c_1 \exp({-\hat\lambda_1 t}) + \hat c_2 \exp({\hat\lambda_2 t})$ in \eqref{eq: C2} with $t\in [t_1^1, t_2^1]$ is convex, we have that $|e(t)| \leq \gamma_1(0)\gamma_2(T_a)\bar E$ for all $t \in [0, t_2^1]$. Recursively, 
this implies that $|e(t)| \leq \gamma_1(0)\gamma_2(T_a)\bar E$ for all $t \geq 0$.
\end{proof}
}


\begin{Remark}
{Consider a set $X_0$, 
{and for a given $x_0\in X_0$ such that  $x(0)=x_0$, define the set $\hat X_0(x_0) :=\{x \in \mathbb{R}^n : x \in x_0+\bar E\mathbb{B} \}$. } Notice that thanks to Theorem \ref{Th:TBoundError}, for 
{each} $x_0\in X_0$, 
and each $\hat x_0\in  \hat X_0(x_0)$ 
we have that each solution pair $t \mapsto \pno{(x(t),\hat x(t))}$ to \eqref{eq: actual system} from $x(0)=x_0, \hat x(0) = \hat x_0$ satisfies 
\begin{itemize}
    \item[1)] Boundedness of error at all times: $|x(t)-\hat x(t)|\leq \bar E$ for all $t\geq 0$\pno{;}
    \item[2)] Maximum error at the beginning of each attack: $|x(t_1^i)-\hat x(t_1^i)|\leq \gamma_1(T_{na})$ for each $i \in \mathbb{N}_{> 0}$ . 
\end{itemize}}
Under \pno{an} attack, it is possible that the error grows, and when there is no attack, \pno{the error} decreases. However, \pno{using} the proposed observer, the norm of the error always remains bounded by $\gamma_1(0)\gamma_2(T_a)\bar E$, \pno{as long as Assumption \ref{ass:bounderror} on the initial estimation error holds}.
\end{Remark}





\pno{\section{Observer-Based Feedback Law Design}}
\NotACC{In this section, we present a set construction process to solve part 1 of Problem $(\star)$ and, based on it,  a control design scheme that solves part 2. First, we define the sets that are going to be used in the control design. }

\pno{\subsection{Construction of Sets of Initial Conditions}}
%
%
Consider a closed set $S  \subset \mathbb{R}^n$, 
$T_a, T_{na}>0$, maps $\gamma_1$ and $ \gamma_2$ as in \eqref{eq: C1} and \eqref{eq: C2}, and $\bar E>0$ in Assumption \ref{ass:bounderror}. 
\pno{Pick} \pn{$\varepsilon >(1+\gamma_1(0)\gamma_2(T_a))\bar E$}
Define the set of initial states as 
\small{
\begin{align}\label{eq: set X0}
 X_0 \coloneqq S \setminus (\partial S + \varepsilon \mathbb B) .
\end{align}
}\normalsize
Note that under Assumption \ref{ass:bounderror}, \pno{$X_0$ is nonempty.} Now, given $x_0\in X_0$, set  $x(0)=x_0$ and define the \pno{set-valued map} 
\small{
\begin{align}\label{eq: set hat X0}
    \hat X_0(x_0) \coloneqq x_0 + 
    \bar E\mathbb{B}.
\end{align}
}\normalsize
Thus, for each $x_0 \in X_0$ and $\hat x_0\in \hat X_0(x_0)$, it holds that $|x_0-\hat x_0|\leq \bar E$. Additionally, notice that {$\hat x_0\in \tilde{X}$}, 
where 
\small{
\begin{align}\label{eq: tilde X}
    \tilde X \coloneqq X_0 + 
    \bar E \mathbb B
\end{align}
}\normalsize
\pno{which is an inflation of $X_0$ by $ 
\bar E$.}
\pno{This} construction of the sets of initial conditions, namely, $X_0$ and $\hat X_0$, \pno{leads to} conditional invariance of $S$, as shown below.

\begin{Lemma}\label{lemma: bounded error}
Given the system \eqref{eq: actual system}, the observer \eqref{eq: hat x switch}, the observer-based feedback law $\kappa$ \eqref{eq: u switch}, a closed set $S \subset \mathbb{R}^n$, 
$X_0$ as in \eqref{eq: set X0}, and $\hat X_0$ as in \eqref{eq: set hat X0}, consider the solution $t \mapsto (x(t),\hat x(t))$ to the resulting closed-loop system from the composition of \eqref{eq: actual system} and \eqref{eq: hat x switch} with $\kappa$ from $x(0)\in X_0$, $\hat x(0)\in \hat X_0(x(0))$ and $T_a, T_{na}, \bar E$,  such that conditions of Theorem \ref{Th:TBoundError} are satisfied. If $S \setminus (\partial S + \pn{(1+\gamma_1(0)\gamma_2(T_a))\bar E\mathbb B})\neq \emptyset$ and $ \hat x(t) \in \tilde X$ for all $t \geq 0$, then $ x(t) \in S$ for all $t \geq 0$. 
\end{Lemma}
\NotACC{
\begin{proof}
 From the definition of the sets $X_0$ and $\tilde X$, and $\varepsilon \geq (1+\gamma_1(0) \gamma_2(T_a)) \bar E $ 
 it follows that 
 $ X_0 \subset S \setminus (\partial S + (1+ \gamma_1(0) \gamma_2(T_a)) \bar E \mathbb B)$ and 
  $\tilde X \subset  S \setminus (\partial S + \gamma_1(0) \gamma_2(T_a)\bar E\mathbb B )$.
 Thus, under the assumption that $\hat x(t) \in \tilde X$ for all $t \geq 0$
with 
$|\hat {x}(t) - x(t)| \leq \gamma_1(0)\gamma_2(T_a)\bar E$ for all $t \geq 0$,
 it follows that
 $x(t) \in S$ for all $t \geq 0$.
\end{proof}
}
In words, the set of initial states $X_0$ and the set of initial estimates $\hat X_0$ are defined such that the initial estimation error is upper bounded by $\bar E$. 
Furthermore, we define $\tilde X$ in \eqref{eq: tilde X} as the set resulting from an inflation of $X_0$ by $
\pn{\bar E}$. Under this construction, 
\pno{
for the resulting closed-loop system from the composition of \eqref{eq: actual system} and \eqref{eq: hat x switch} with $\kappa$,}
forward invariance of $\tilde X$ for the observer \eqref{eq: hat x switch} implies conditional invariance of the set $S$ for the system \eqref{eq: actual system}
with respect to $X_0$.
%
Thus, the control objective is to enforce {the estimate $\hat x$ in the set $\tilde X$} at all times to guarantee safety of $S$.

\subsection{QP-based Feedback Law Synthesis}
We use a control barrier function (CBF)-based approach for guaranteeing forward invariance of a subset $\bar X$ of the set $\tilde X$ in \eqref{eq: tilde X} for \eqref{eq: hat x switch} (see \cite{ames2017control}). In order to use CBF for forward invariance, we need a zero sublevel set representation of the set $\bar X$. To this end, consider the function $h:\mathbb R^n\rightarrow\mathbb R$ and define a set
{\small
\begin{align}\label{eq: bar X}
    \bar X \coloneqq \{\hat x\; |\; h(\hat x)\leq 0\}\subset {\tilde X}.
\end{align}
}

Given an observer-based feedback law $\kappa$ assigning the input $u = \kappa(t, \hat x,\bar y)$ of \eqref{eq: hat x switch}, consider a solution $t \mapsto \hat x(t)$ to \eqref{eq: hat x switch} from $\hat x(0)\in \bar X$. \pno{For the given measurement $\bar y$,} it is sufficient to ensure that for each $\hat x(0)\in \bar X$, 
the estimate satisfies $\hat x(t)\in \bar X \subset \tilde X$, for all $t\geq 0$. The CBF condition for guaranteeing this 
when there is no attack is:
\small{
\begin{align}\label{eq: CBFCondnA}
    \frac{\partial}{\partial \hat x}h(\hat x(t))& \left(A\hat x(t)+ B\kappa_1(\hat x(t), \bar y(t))\right. \nonumber \\
    & \left.+ L(\bar y(t)-C\hat x(t))\right)\leq \alpha_1(-h(\hat x(t))), 
\end{align}}\normalsize
for all $t\geq 0$, where $t \mapsto \bar y(t)$ is the measured output signal, and the CBF condition under attack is
\small{
\begin{align}\label{eq: CBFConduA}
    \frac{\partial}{\partial \hat x}h(\hat x(t))& \left(A\hat x(t)+ B\kappa_2(\hat x(t), \bar y(t))\right. \nonumber \\
    & \left.+ \tilde L(\bar y_s(t)-\tilde C\hat x(t))\right)\leq \alpha_1(-h(\hat x(t))), 
\end{align}}\normalsize 
for all $t\geq 0$, where $t \mapsto y_s(t)$ is the secured output signal and $\alpha_1, \alpha_2$ are class-$\mathcal K$ functions. 
We can use a Quadratic Programming (QP) formulation to compute the input $u$ in the respective cases. 

Consider the following QP for each $\hat x\in {\bar X}$ {and $\bar y$ such that $x \in S$} for input synthesis when there is no attack:
\small{
\begin{subequations}\label{QP no attack}
\begin{align}
\hspace{10pt}\min_{(v, \eta)} \quad \frac{1}{2}|v-K\hat x|^2 + & \frac{1}{2}\eta^2\\
     \textrm{s.t.} \;\quad \; \; 
    \frac{\partial}{\partial \hat x}h(\hat x)\left(A\hat x+ Bv + L(\bar y-C\hat x)\right) \leq & -\eta h(\hat x),
\end{align}
\end{subequations}}\normalsize
where $K$ is the optimal LQR gain for the pair $(A,B)$. 
Next, we use a similar QP to compute the input under attack. Consider the following QP for each $\hat x \in {\bar X}$ {and $y_s = \tilde C x$ such that $x \in S$}:
\small{
\begin{subequations}\label{QP under attack}
\begin{align}
\hspace{10pt}\min_{(v_s, \zeta)} \quad \frac{1}{2}|v_s-K\hat x|^2 + & \frac{1}{2}\zeta^2\\
     \textrm{s.t.} \;\quad \; \; 
    \frac{\partial}{\partial \hat x}h(\hat x)\left(A\hat x+ Bv_s + {\tilde L}(y_s-\tilde C\hat x)\right) \leq & -\zeta h(\hat x).
\end{align}
\end{subequations}}\normalsize  
{The objective functions in \eqref{QP no attack} and \eqref{QP under attack} set the convex minimization problem to obtain the closest control action to the LQR control that satisfies the constraints. The additional decision variables, namely $(\eta, \zeta)$, respectively, are slack variables.}
Denote the solutions to \eqref{QP no attack} and \eqref{QP under attack} as {$t \mapsto u_1^*(\hat x(t),\bar y(t))$ and $t \mapsto u_2^*(\hat x(t), \bar y(t))$}, respectively. \pno{To} guarantee continuity of these solutions with respect to $\hat x$, we need to impose the strict complementary slackness condition (see \cite{garg2022fixed}). In brief, if the $i-$th constraint of \eqref{QP no attack} (or \eqref{QP under attack}), with $i \in \{1, 2\}$, is written as $G_i(\hat x,\bar y,u_{QP})\leq 0$ with $u_{QP} = (v,\eta)$ (respectively, $u_{QP} = (v_s, \zeta)$ for \eqref{QP under attack}), and the corresponding Lagrange multiplier is $\bar\lambda_i\in \mathbb R_{\geq 0}$, then strict complementary slackness requires that $\bar\lambda_i^*G(\hat x,\bar y,u_{QP}^*)<0$, 
where $u_{QP}^*$ and $\bar\lambda_i^*$ denote the optimal solution and the corresponding optimal Lagrange multiplier, respectively. We are now ready to state the second main result of the paper. 

\begin{Theorem}\label{TH: 2}
\pno{Given system \eqref{eq: actual system},
suppose that Assumptions \ref{assum A B cont} and \ref{ass:bounderror} hold.}
For the attack model \eqref{eq: attack model}, the observer \eqref{eq: hat x switch}, and a closed set $S\subset \mathbb{R}^n$, 
let $X_0$, $\hat X_0$, $\tilde X$ and $\bar X$ be given as in \eqref{eq: set X0}-\eqref{eq: bar X}, and assume that the strict complementary slackness holds for the QPs \eqref{QP no attack} and \eqref{QP under attack} for all $\hat x\in {\tilde X}$.
The following holds\pno{:}
\begin{enumerate}
\item If $S \setminus (\partial S + \pn{(1+}\gamma_1(0)\gamma_2(T_a))\bar E\mathbb B)\neq \emptyset$, then, for each $\hat x \in \textnormal{int}( \bar X)$, the QPs \eqref{QP no attack} and \eqref{QP under attack} are feasible and their respective solutions $t \mapsto u_1^*(\hat x(t), \bar y(t)), t \mapsto u_2^*(\hat x(t), \bar y(t))$ are continuous on $\textnormal{int}(\bar X)$. 

\item For each $x_0\in X_0$ and $\hat x_0\in \bar X \cap \hat X_0(x_0)$, 
each solution pair $t \mapsto (x(t), \hat x(t))$ to the closed-loop system resulting from assigning the input $u$ of \eqref{eq: actual system} and \eqref{eq: hat x switch} to the {observer-based feedback law} $\kappa$ in \eqref{eq: u switch} with $\kappa_1(\hat {x}, \bar y) = u_1^*(\hat x, \bar y)$ and $\kappa_2(\hat {x}, \bar y) = u_2^*(\hat x, \bar y)$,
 satisfies
$\hat x(t)\in \bar X$ 
and $x(t) \in S$ 
for all $t\geq 0$.
\end{enumerate}
\end{Theorem}
\NotACC{
\begin{proof}
Feasibility and continuity of the solutions of the QPs \eqref{QP no attack} and \eqref{QP under attack} follow from \cite[Lemma 5]{garg2022fixed} and \cite[Theorem 1]{garg2022fixed}, respectively. 
\pn{From feasibility of the QPs and continuity of its solutions, there exists a continuous control input $u$ such that the CBF condition \eqref{eq: CBFConduA} holds along the closed-loop trajectory $\hat x(t)$.
Thus, it follows that 
 the set $\tilde X$ is forward invariant for the observer \eqref{eq: hat x switch}, and hence}
$\hat x(t)\in \tilde X$ for all $t\geq 0$ and for each $\hat x(0)\in \tilde X_0$. 
Thanks to Theorem \ref{Th:TBoundError}, for each $x(0)\in X_0$, one has $|e(t)|\leq \gamma_1(0)\gamma_2(T_a)\bar E$, which from Lemma \ref{lemma: bounded error} implies $x(t)\in S$ for all $t\geq 0$. 
\end{proof}
}

\NotACC{Thus, the proposed observer-based feedback framework, based on a \pno{switching} observer and \pno{a switching control scheme}, can keep the system safe even under output attacks. Next, we evaluate our proposed scheme via numerical experiments.}

\section{Numerical Example}
Consider a system $\mathcal S $ as in \eqref{eq: actual system}, with state $x=(x_1,x_2) \in \mathbb{R}^2$, input $u \in \mathbb{R}$, and 
dynamics
$\dot x = (x_2, u), 
y = (x_1, x_2)$ 
where 
\pno{$y_a=x_1$} is only available when there are no attacks. 
DoS attacks have maximum duration of $T_a=1.6$ seconds and are launched only after at least $T_{na}=0.047$ seconds without an attack. Here, $u$ is designed such that every response $t \mapsto x(t)$ to 
$\mathcal{S}$
satisfies $ x(t) \in S:=\{(x_1,x_2) \in \mathbb{R}^2 : x_1^2 + 2 x_2^2+2 x_1 x_2-35\leq 0\}$ for all $t\geq 0$, given that $x(0) \in X_0:= S \setminus (\partial S + \varepsilon \mathbb B)$, with {$\varepsilon=2.01$}.

An observer as in \eqref{eq: hat x switch} is designed. Given that Assumption \ref{assum A B cont}
is satisfied, and by setting $L=\left[\begin{smallmatrix} 
32 && 0.5 
\\ 
0.5 && 32
\end{smallmatrix}\right]$ 
and 
$\tilde L =\left[\begin{smallmatrix} 0.05 \\ 3.2 \end{smallmatrix}\right] $, we have $\lambda(A- L  C)=-31.75\pm\textit{i}0.43$, 
and
$\lambda(A-\tilde L \tilde C)=\pno{\{0.5,-3.2\}}$. 
Given $x_0=(5.3,-2.4)$, $\hat x_0=(4.9,-2.1)$, and {$\bar E=0.55$}, we have that $|e(0)|=0.5 \leq \bar E$, so Assumption \ref{ass:bounderror} holds.

{Thus, by applying Lemma \ref{Lemma: Bound un attack}, with 
$P=\left[\begin{smallmatrix} 1 && 0 \\ 0 && 1 \end{smallmatrix}\right], 
Q=\left[\begin{smallmatrix}
63 && 0 
\\ 
0 && 64
\end{smallmatrix}\right] $, 
and given that every pair of subsequent attacks are separated by at least $T_{na}$ seconds, the estimation error satisfies 
$|e(t)|\leq \gamma_1(t-t^{i}_2) e(t^{i}_2)$ for all $t \in [t^{i}_2,t_1^{i+1}]$, $i\in \mathbb{N}$,  $\gamma_1( T_{na}=0.047)=
    0.226,$ and is displayed in green\footnote{Code at 
    https://github.com/HybridSystemsLab/SafeRecovery-DoSAttacks} 
    in Figure \ref{fig:statesex}}.
 Given that the growth rate of the exponential defining the function $\gamma_1$ is negative, 
 {the bound on the error norm} decreases at each interval without attacks.
%


\begin{figure}[t]
\hspace{-0.5cm}
\includegraphics[width=1.05\columnwidth,clip]{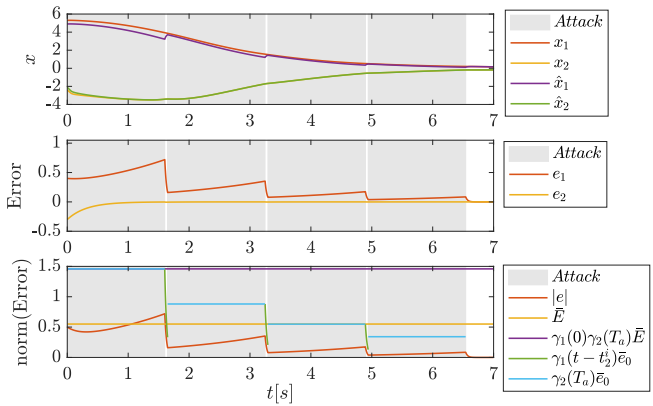}
	\caption{Solutions to the 2D system and state estimation error during worst-case attacks of  $T_a=1.6s$, for $x_0=(5.3,-2.4)$, $\hat x_0=(4.9,-2.1)$, and  $\bar E=0.55$. {In the third plot, the bound (purple) is defined as in Theorem 1.}}
	\label{fig:statesex}
\end{figure}

{In addition, by applying Lemma \ref{Lemma: Bound u attack} 
with 
$\hat c_1
=1.12,
\hat c_2 
=1.19, 
\hat \lambda_1=3.2,
{\hat \lambda_2
=0.5,} \> 
\hat P=\left[\begin{smallmatrix}
1 && 0
\\ 
0 && 1 
\end{smallmatrix}\right], 
\hat Q=\left[\begin{smallmatrix} 6.4 && 0 
\\ 
0 && 6.4 
\end{smallmatrix}\right] $, 
$\Phi=\left[\begin{smallmatrix} 
1 && -0.25 
\\ 0 && 0.97 \end{smallmatrix}\right]$, and
{$\hat A_{22}= 0.5$,}
 given that every attack has a maximum duration of $T_{a}$ seconds, the estimation error satisfies 
$|e(t)|\leq \gamma_2(T_a)|e(t_1^i)| $ for all $t \in [t_1^i,t^i_2]$, $i\in \mathbb{N}_{>0}$ 
where  $\gamma_2( T_{a})
=2.65,$ and is displayed in light blue in Figure \ref{fig:statesex}.}
 %
%
%
\pn{Thanks to Theorem 1, given that $\gamma_1(T_{na})\gamma_2(T_1) \leq 1$, the error satisfies $|e(t)|\leq c_1 \gamma_2(T_a)\bar E = 1.46 $ for all $t\geq 0$.}

In Figure \ref{fig:ppex}, the set $X_0$ is a deflation of the set $S$ by $\varepsilon$, and the set $\tilde X$ is an inflation of the set $X_0$ by $\bar E$. The set of initial estimations, $\hat X_0(x_0)$, is defined as the ball of radius $\bar E$ centered at $x_0$.
Thus, the estimator $\hat x$ is initialized at $X_0(x_0)\subset \tilde X$.  
The set $\bar X :=\{(x_1,x_2) \in \mathbb{R}^2 : h(x)\leq 0\}\subset  \tilde X$ is defined by the barrier function $h(x)=x_1^2 + 2 x_2^2+2 x_1 x_2-12.5$.
Given that {the set $S\subset \mathbb R^n$ is such that $S \setminus (\partial S + \pn{(1+\gamma_1(0)\gamma_2(T_a))\bar E\mathbb B})\neq \emptyset$}, by assigning $K=[2.3016,\>\>    2.3671]$ and solving the QPs \eqref{QP no attack} and \eqref{QP under attack} at every point of the trajectory $\hat x(t) \in \bar X$ to assign the input action, 
thanks to Theorem \ref{TH: 2}, we ensure that $\hat x(t) \in \bar X$ for all $t$, and consequently, $x(t) \in S$ for all $t$. 

\begin{figure}[t]
	\hspace{-0.7cm}
	\includegraphics[width=1.14\columnwidth,clip]{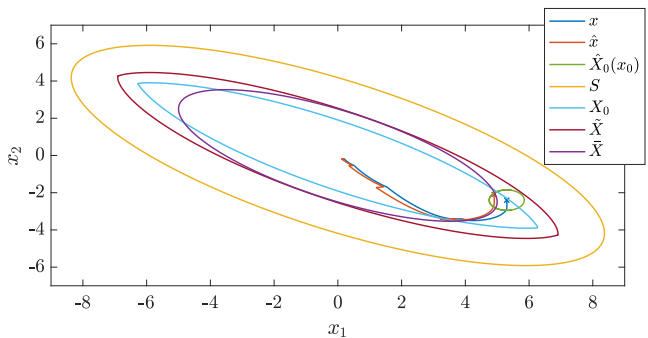}
	\caption{Phase portrait of $\dot x = (x_2, u), 
y = (x_1, x_2)$ with state estimation for safe recovery of DoS attacks in the measurements of $x_1$. 
By initializing the estimation $\hat x$ in the $\bar E-$ball (green) around $x(0)$, the set $\bar X$ (purple) is rendered forward invariant for $\hat x$ (orange), and the safe set $S$ (yellow) conditionally invariant for $x$ (dark blue) with respect to the set of allowed initial states, namely $X_0$ (light blue), via the control barrier function.
The set $\tilde X$ (scarlet) denotes the allowed initial \pno{observer} states.
	}
	\label{fig:ppex}
\end{figure}

\NotACC{\subsection{Example 2: 10-D System}
To show the potential of scalability of our design, consider a system with state $x \in \mathbb R^n$, input $u \in \mathbb{R}^m$, and dynamics as in \eqref{eq: actual system}, with randomly generated parameters $A \in \mathbb R^ {n \times n}, B \in \mathbb R^ {n \times m},$ and $C \in \mathbb R^ {p \times n}$, where $n=10, m=5, p=5$, satisfying
$\textup{rank} (\tilde{\mathcal O} (C,A))=n, \textup{rank} (\tilde{\mathcal C} (A,B))=n, \textup{rank} (\tilde{\mathcal O} (\tilde C,A))=n-\tilde p$, namely, $\tilde p$ randomly chosen outputs are attacked.

DoS attacks have maximum duration of $T_a=\pn{0.2}$ seconds and are launched only after at least $T_{na}=\pn{18.9}$ seconds without an attack.

Here, $u$ is designed such that every response $t \mapsto x(t)$ to 
$\mathcal{S}$ 
satisfies $x(t) \in S:=\{x \in \mathbb{R}^{10} : x_1^2 + 2 x_2^2 + x_3^2 + x_4^2 + x_5^2 + x_6^2 + x_7^2 + x_8^2 + x_9^2 + x_{10}^2 + 2 x_1 x_2-25\leq 0\}$ for all $t\geq 0$, given that $x(0) \in X_0:= S \setminus (\partial S + \varepsilon \mathbb B)$, with $\varepsilon=\pn{7}$.

An observer as in \eqref{eq: hat x switch} is designed. Given that Assumption \ref{assum A B cont}
is satisfied, and by optimally calculating $L$ as the gain of the action $u=-L^\top x$ that minimizes the quadratic cost function $J(u)=\int_0^\infty (x^\top 100I_{n \times n}x+u^\top 0.01 I_{p\times p} u) dt$ subject to the dynamics $\dot x = A^\top x+ C^\top u$, 
and 
$\tilde L $ as the corresponding values to $L$ after deleting the attacked signals, we have $\lambda_m(A- L  C)=-371.1$, 
$\lambda_M(A- L  C)=-1.61 \pm 1.36\textit{i} $, 
$\lambda_m(A-\tilde L \tilde C)=-332.2$, and $\lambda_M(A-\tilde L \tilde C)=1.91$. 

%

\begin{figure}[t]
	\centering
	\includegraphics[width=0.8\columnwidth,clip]{AttackPlots2.png}
	\vspace{-1.3cm}
	\caption{Solutions to a larger scale system and error of the state estimator with respect to the actual state during \pn{longest possible}  attacks and no attacks intervals for  $T_a=0.2, T_{na}=18.9, x_0=(2, -1, 1, 1, 1, 1, 1, 1, 1, 1)$, $\hat x_0=x_0+\textup{rand}(n,1)$, and $\bar E=3.5$. In the third plot, the bound in yellow is defined as in Theorem 1. 
	}
	\label{fig:statesex2}
\end{figure}

%

The estimator $\hat x$ is initialized at $X_0(x_0)\subset \tilde X$.  
The set $\bar X :=\{(x \in \mathbb{R}^10 : h(x)\leq 0\}\subset  \tilde X$ is defined by the barrier function $h(x)=x_1^2 + 2 x_2^2 + x_3^2 + x_4^2 + x_5^2 + x_6^2 + x_7^2 + x_8^2 + x_9^2 + x_{10}^2 + 2 x_1 x_2-17$.
\pn{Given that $\varepsilon \geq \bar E (\gamma_1(0) \gamma_2(T_a) +1) $, by solving the QPs \eqref{QP no attack} and \eqref{QP under attack} at every point of the trajectory $\hat x(t) \in \bar X$, 
we show that thanks to Theorem \ref{TH: 2}, by rendering $\hat x(t) \in \bar X$ for all $t$, we guarantee $x(t) \in S$ for all $t$.}
}{}
\section{Conclusion and Future Work}
In this paper, we present a switched controller design that, together with a switched observer, ensures a linear time-invariant system to recover safely from finite-time DoS attacks in some of the system outputs. Conditional invariance of a set is guaranteed with respect to a subset of initial conditions by employing a barrier function approach and bounding the estimation error at all times. 
Future works include studying safe-recovery controllers under uncertainty in the model parameters, noise in the unattacked sensors, nonlinearities in the system dynamics, and only approximate information on the attack times.
In addition, an implementation of a finite-time observer and a tighter bound to relax the conservatism of the conditions are to be considered.
\bibliographystyle{IEEEtran}
\bibliography{myreferences}

\begin{thebibliography}{10}
\providecommand{\url}[1]{#1}
\csname url@samestyle\endcsname
\providecommand{\newblock}{\relax}
\providecommand{\bibinfo}[2]{#2}
\providecommand{\BIBentrySTDinterwordspacing}{\spaceskip=0pt\relax}
\providecommand{\BIBentryALTinterwordstretchfactor}{4}
\providecommand{\BIBentryALTinterwordspacing}{\spaceskip=\fontdimen2\font plus
\BIBentryALTinterwordstretchfactor\fontdimen3\font minus
  \fontdimen4\font\relax}
\providecommand{\BIBforeignlanguage}[2]{{%
\expandafter\ifx\csname l@#1\endcsname\relax
\typeout{** WARNING: IEEEtran.bst: No hyphenation pattern has been}%
\typeout{** loaded for the language `#1'. Using the pattern for}%
\typeout{** the default language instead.}%
\else
\language=\csname l@#1\endcsname
\fi
#2}}
\providecommand{\BIBdecl}{\relax}
\BIBdecl

\bibitem{chong2019tutorial}
M.~S. Chong, H.~Sandberg, and A.~M. Teixeira, ``A tutorial introduction to
  security and privacy for cyber-physical systems,'' in \emph{2019 18th
  European Control Conference (ECC)}.\hskip 1em plus 0.5em minus 0.4em\relax
  IEEE, 2019, pp. 968--978.

\bibitem{teixeira2015secure}
A.~Teixeira, I.~Shames, H.~Sandberg, and K.~H. Johansson, ``A secure control
  framework for resource-limited adversaries,'' \emph{Automatica}, vol.~51, pp.
  135--148, 2015.

\bibitem{cardenas2008secure}
A.~A. Cardenas, S.~Amin, and S.~Sastry, ``Secure control: Towards survivable
  cyber-physical systems,'' in \emph{2008 The 28th International Conference on
  Distributed Computing Systems Workshops}.\hskip 1em plus 0.5em minus
  0.4em\relax IEEE, 2008, pp. 495--500.

\bibitem{Amin2009DoSNetControl}
S.~Amin, A.~A. C{\'a}rdenas, and S.~S. Sastry, ``Safe and secure networked
  control systems under denial-of-service attacks,'' in \emph{Proceedings of
  the 12th International Conference on Hybrid Systems: Computation and
  Control}, vol. 5469.\hskip 1em plus 0.5em minus 0.4em\relax Springer, Berlin,
  Heidelberg, 2009, pp. 31--45.

\bibitem{krotofil2014cps}
M.~Krotofil, A.~A. C{\'a}rdenas, B.~Manning, and J.~Larsen, ``Cps: Driving
  cyber-physical systems to unsafe operating conditions by timing dos attacks
  on sensor signals,'' in \emph{Proceedings of the 30th Annual Computer
  Security Applications Conference}, 2014, pp. 146--155.

\bibitem{fawzi2014secure}
H.~Fawzi, P.~Tabuada, and S.~Diggavi, ``Secure estimation and control for
  cyber-physical systems under adversarial attacks,'' \emph{IEEE Transactions
  on Automatic control}, vol.~59, no.~6, pp. 1454--1467, 2014.

\bibitem{leblanc2013resilient}
H.~J. LeBlanc, H.~Zhang, X.~Koutsoukos, and S.~Sundaram, ``Resilient asymptotic
  consensus in robust networks,'' \emph{IEEE Journal on Selected Areas in
  Communications}, vol.~31, no.~4, pp. 766--781, 2013.

\bibitem{kafash2018constraining}
S.~H. Kafash, J.~Giraldo, C.~Murguia, A.~A. Cardenas, and J.~Ruths,
  ``Constraining attacker capabilities through actuator saturation,'' in
  \emph{2018 Annual American Control Conference (ACC)}.\hskip 1em plus 0.5em
  minus 0.4em\relax IEEE, 2018, pp. 986--991.

\bibitem{yan2017resilient}
Y.~Yan, P.~Antsaklis, and V.~Gupta, ``A resilient design for cyber physical
  systems under attack,'' in \emph{2017 American Control Conference
  (ACC)}.\hskip 1em plus 0.5em minus 0.4em\relax IEEE, 2017, pp. 4418--4423.

\bibitem{bai2017kalman}
C.-Z. Bai, V.~Gupta, and F.~Pasqualetti, ``On kalman filtering with compromised
  sensors: Attack stealthiness and performance bounds,'' \emph{IEEE
  Transactions on Automatic Control}, vol.~62, no.~12, pp. 6641--6648, 2017.

\bibitem{ames2014control}
A.~D. Ames, J.~W. Grizzle, and P.~Tabuada, ``Control barrier function based
  quadratic programs with application to adaptive cruise control,'' in
  \emph{53rd Conference on Decision and Control}.\hskip 1em plus 0.5em minus
  0.4em\relax IEEE, 2014, pp. 6271--6278.

\bibitem{clark2020control}
A.~Clark, Z.~Li, and H.~Zhang, ``Control barrier functions for safe cps under
  sensor faults and attacks,'' in \emph{2020 59th IEEE Conference on Decision
  and Control (CDC)}.\hskip 1em plus 0.5em minus 0.4em\relax IEEE, 2020, pp.
  796--803.

\bibitem{amin2009safe}
S.~Amin, A.~A. C{\'a}rdenas, and S.~S. Sastry, ``Safe and secure networked
  control systems under denial-of-service attacks,'' in \emph{International
  Workshop on Hybrid Systems: Computation and Control}.\hskip 1em plus 0.5em
  minus 0.4em\relax Springer, 2009, pp. 31--45.

\bibitem{198}
M.~Maghenem and R.~G. Sanfelice, ``Characterizations of safety and conditional
  invariance in dynamical systems,'' in \emph{Proceedings of the American
  Control Conference}, July 2019, pp. 5039--5044.

\bibitem{ames2017control}
A.~D. Ames, X.~Xu, J.~W. Grizzle, and P.~Tabuada, ``Control barrier function
  based quadratic programs for safety critical systems,'' \emph{IEEE
  Transactions on Automatic Control}, vol.~62, no.~8, pp. 3861--3876, 2017.

\bibitem{shen2008semi}
Y.~Shen and X.~Xia, ``Semi-global finite-time observers for nonlinear
  systems,'' \emph{Automatica}, vol.~44, no.~12, pp. 3152--3156, 2008.

\bibitem{garg2022fixed}
K.~Garg, E.~Arabi, and D.~Panagou, ``Fixed-time control under spatiotemporal
  and input constraints: A quadratic programming based approach,''
  \emph{Automatica}, vol. 141, p. 110314, 2022.

\end{thebibliography}


\end{document}